\PassOptionsToPackage{table}{xcolor}
\documentclass[10pt, conference]{IEEEtran}
\usepackage{amsmath} 
\usepackage{amsthm} 
\usepackage{amssymb} 
\usepackage{graphicx} 
\usepackage{multicol} 
\usepackage{sparklines} 
\usepackage{subcaption}
\usepackage{cleveref}

\let\svthefootnote\thefootnote
\newcommand\blankfootnote[1]{%
  \let\thefootnote\relax\footnotetext{#1}%
  \let\thefootnote\svthefootnote%
}

\begin{document}
	
\title{Research, Develop, Deploy: Building a Full Spectrum Software Engineering and Research Department}
\author{\IEEEauthorblockN{Reed Milewicz, James Willenbring, and Dena Vigil}
\IEEEauthorblockA{Department of Software Engineering and Research\\Sandia National Laboratories, 1611 Innovation Pkwy SE, Albuquerque, New Mexico 87123}}
\maketitle

\begin{abstract}
        
At Sandia National Laboratories, the Software Engineering and Research Department seeks to provide sustainable career pathways for research software engineers (RSEs). The conceptual model for our organization follows what we call a Research, Develop, and Deploy (RDD) workflow pattern, enabling RSEs to partner with research and deployment specialists. We argue that this interdisciplinary model allows our department to act as an incubator and an accelerator for impactful ideas. We describe these tactics and our experiences as a RSE team in a scientific computing center.

\end{abstract}


\blankfootnote{Sandia National Laboratories is a multimission laboratory managed and operated by National Technology \& Engineering Solutions of Sandia, LLC, a wholly owned subsidiary of Honeywell International Inc., for the U.S. Department of Energy's National Nuclear Security Administration under contract DE-NA0003525. SAND2020-11072 C. Presented at Research Software Engineers in HPC (RSE-HPC-2020), co-located with SC'20.
}

\section{Introduction}


The past decade has seen a dramatic growth in demand for scientific computing, and there is a well-recognized need for professionals who can advance software engineering practice within the scientific domain. To meet these workforce requirements, we need organizational structures that give recognition to software-focused personnel and provide pathways for career advancement~\cite{baxter2012research}\cite{cohen2020four}. On the institutional front, many research organizations have made strides towards creating positions and groups for RSEs. The recently formed Department of Software Engineering and Research has established a RSE team to support the Center for Computing Research at Sandia National Laboratories~\cite{ccrwebsite}. Key to our strategy is integrating RSEs into a cross-cutting R\&D team alongside personnel with complementary skillsets, namely software engineering research, DevOps, and IT service management. In this paper, we describe our group's structure and strategy and reflect on our experiences. 

\section{Background}

The Department of Software Engineering and Research is an R\&D organization responsible for advancing the study and practice of software engineering to support scientific software. Formed in late 2018, this group was the culmination of several years of effort to unite software-focused professionals scattered across numerous departments within the center. This new department was preceded by the Software Engineering, Maintenance, and Support (SEMS) project, which provided a network for RSEs to collaborate with like-minded peers; by working collectively, SEMS was able to provide tools, training, and support to computational scientists and their projects~\cite{willenbring2020}. However, SEMS was not an official department, and this made it difficult to hire, retain, and provide a career path for RSEs. Different SEMS team members answered to different managers -- none of whom had software engineering as a primary focus -- and the structures for incentives and career advancement in these research departments were geared towards domain scientists and mathematicians. Instituting official organizations for RSEs, ones that recognize their contributions and create career pathways for them, is critical to support workforce needs; this is why our department was created.

Our department is a matrix-managed team that provides flexible, on-demand staffing for development, consultation, and support to other departments within the Center for Computing Research. In this way, it shares characteristics in common with analogous RSE organizations elsewhere. As the name ``Software Engineering \textit{and} Research'' suggests, our team is an R\&D department that features both software engineering research and research software engineering. This is comparable to the Research Software and Data Science (RSDS) group at the University of Manchester, which also features both an engineering component and an applied science component~\cite{katz2019research}. In addition to this, our work also encompasses DevOps and ITSM activities in that we offer community infrastructure and helpdesk support. Perhaps the closest fit would be the Research Software Engineering Group at Oak Ridge National Laboratories, which provides embedded software development and project management support as well as contributing to and leading research projects in computational science and software engineering~\cite{billings2020}.

\subsection{Organizational Model}


Our team structure follows what we call a Research, Develop, and Deploy (RDD) workflow pattern, centered around three principal workflows:

\begin{itemize}

        \item \textbf{Research}: Our team conducts fundamental and applied research in software engineering; specialists in this focus area tend to have a PhD in computer science or a domain science and a passion for software engineering. Given the importance of our work for the nation, we have a reponsibility to act on the basis of the best available evidence. Without sufficient understanding, we run the risk of solving the wrong problems, having the ``right'' solutions go unused, or failing to seize emerging opportunities. A key way in which we acquire and retain that understanding is through rigorous, systematic investigation -- that is, research (see ~\cite{milewicz2020ebp}). For that reason, we actively particpate in the software engineering research community to bring the best ideas to Sandia through evidence-based practice and to create new knowledge for improving scientific software. 
        
        \item \textbf{Develop}: A key element of our work is in teaming with application and algorithm researchers to provide embedded development, maintenance, and support; team members who focus on development work tend to be RSEs of various stripes, ranging from computer science graduates to staff with a background in science or mathematics who have transitioned into a software-focused career. One important type of software development service provided by our department is scientific programming. This type of development requires scientific domain-specific knowledge and skills, such as a familiarity with domain algorithms and experience using MPI, OpenMP, and/or CUDA. Our team also provides software development services for core software engineering needs, such as testing, build systems, and other areas that are not specific to the HPC domain. As part of our outreach, our team also regularly offers training to researchers on useful topics like Git workflows.

        \item \textbf{Deploy}: We firmly believe that robust, scalable, and sustainable infrastructure for software projects is vital to the scientific computing mission of our center; for this reason, we have a contingent of staff who focus on DevOps and/or IT service management to lead these efforts. Case in point: we maintain Jenkins-based build/test farms, a common dependency management system, off-the-shelf tools like Jira and Confluence, and tailored infrastructure solutions for projects. 
        

\end{itemize}

\begin{figure}[h]
        \centering
	\includegraphics[height=5cm]{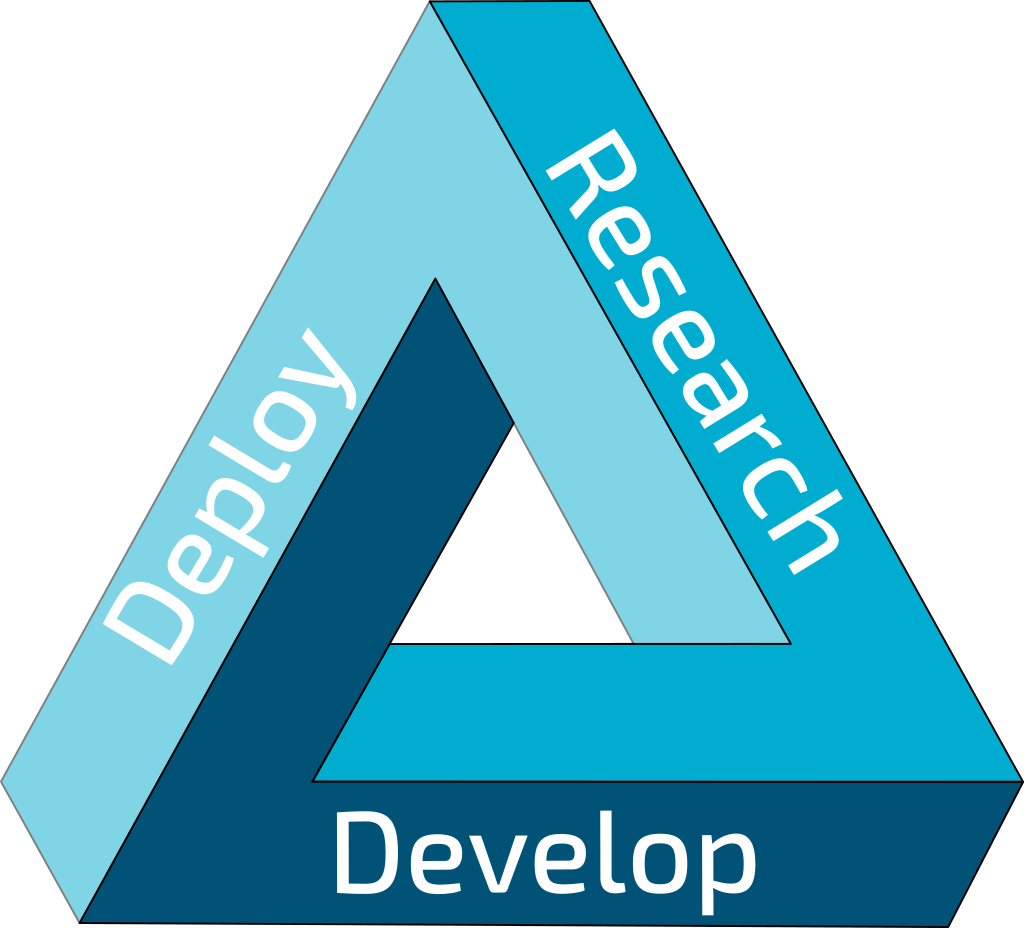} 
	\caption[Department Strategy]{A visual metaphor for our organization's strategic posture. We aim to cultivate a critical mass of staff in each capability area, each of which mutually reinforces the other.}
\end{figure}

Currently, our team is composed of around two dozen professionals -- the majority of whom are RSEs -- who participate in one or more of these workflows. While our RSEs focus primarily on embedded development work with scientific software projects, they also participate in our team's research and deployment activities that amplify their impact. Meanwhile, their peer professionals in research and deployment engage with them regularly to ensure our team's products and services can meet their customers' needs. 



\section{Impact}

Just like RSE groups at other institutions, we continue to grow and learn, and we are not claiming to have a definitive answer as to how best to structure a RSE organization. That being said, we believe that placing RSEs into an interdisciplinary team makes our team more cohesive, creative, and productive. 

First, these roles are tightly integrated: our research work aims to create new knowledge to assist development and deployment activities, our development work keeps research grounded in real-world needs and informs deployment activities, and our deployment work drives research priorities in tooling and support. We believe that it is important to have a critical mass of skills in each workflow in order to have a holistic picture of how to best prioritize, plan, and execute activities.

Second, as an R\&D department, we are expected not just to provide software engineering as a service, but also to be innovators in tools and techniques for scientific programming. Having a research component both legitimizes our work within a research center and creates new opportunities for RSEs to build their careers. Not only are our RSEs involved in research literature reviews and experiments with our stakeholders (cf. \cite{jay2017identifying}), they also publish their own white papers and experience reports at workshops and conferences. Meanwhile, we have a long-running round-table series on best practices in software engineering where team members come together to deliberate and discuss the processes and principles that lead to high-quality software. Activities like these help our team members grow as first-class professionals in our center.

Finally, having a workforce that spans the entire R\&D lifecycle from early research to customer support makes us far more productive than if we solely specialized in embedded development work. When our RSEs encounter unknowns or hit a wall, they can engage with our research staff to find answers, and they are able to leverage our DevOps/ITSM infrastructure to track work items on Jira, execute tests on Jenkins, or handle tickets on our helpdesk. We ensure that whenever a RSE is assigned to a customer project, they are never going alone.

\section{Conclusion}

In this position paper, we described the Department of Software Engineering and Research at Sandia National Labs, and its practice of placing RSEs alongside software research and deployment personnel to increase their impact and support their careers. We hope that our work adds to the discussion on how to build institutional structures for RSEs.

\bibliographystyle{IEEEtran}
\bibliography{rsehpc}

\begin{thebibliography}{1}
\providecommand{\url}[1]{#1}
\csname url@samestyle\endcsname
\providecommand{\newblock}{\relax}
\providecommand{\bibinfo}[2]{#2}
\providecommand{\BIBentrySTDinterwordspacing}{\spaceskip=0pt\relax}
\providecommand{\BIBentryALTinterwordstretchfactor}{4}
\providecommand{\BIBentryALTinterwordspacing}{\spaceskip=\fontdimen2\font plus
\BIBentryALTinterwordstretchfactor\fontdimen3\font minus
  \fontdimen4\font\relax}
\providecommand{\BIBforeignlanguage}[2]{{%
\expandafter\ifx\csname l@#1\endcsname\relax
\typeout{** WARNING: IEEEtran.bst: No hyphenation pattern has been}%
\typeout{** loaded for the language `#1'. Using the pattern for}%
\typeout{** the default language instead.}%
\else
\language=\csname l@#1\endcsname
\fi
#2}}
\providecommand{\BIBdecl}{\relax}
\BIBdecl

\bibitem{baxter2012research}
R.~Baxter, N.~C. Hong, D.~Gorissen, J.~Hetherington, and I.~Todorov, ``{The
  Research Software Engineer},'' in \emph{Digital Research Conference, Oxford},
  2012, pp. 1--3.

\bibitem{cohen2020four}
J.~Cohen, D.~S. Katz, M.~Barker, N.~P.~C. Hong, R.~Haines, and C.~Jay, ``{The
  Four Pillars of Research Software Engineering},'' \emph{IEEE Software}, 2020.

\bibitem{ccrwebsite}
``{Center for Computing Research},'' \url{https://www.cs.sandia.gov/},
  accessed: 2020-10-01.

\bibitem{willenbring2020}
J.~Willenbring and R.~Milewicz, ``{Moving Forward Together: How a Software
  Engineering Department Can Impact Developer Productivity in a Research
  Organization},'' \emph{2020 Collegeville Workshop on Scientific Software},
  2020.

\bibitem{katz2019research}
D.~S. Katz, K.~McHenry, C.~Reinking, and R.~Haines, ``{Research software
  development \& management in universities: case studies from Manchester's
  RSDS group, Illinois' NCSA, and Notre Dame's CRC},'' in \emph{2019 IEEE/ACM
  14th International Workshop on Software Engineering for Science
  (SE4Science)}.\hskip 1em plus 0.5em minus 0.4em\relax IEEE, 2019, pp. 17--24.

\bibitem{billings2020}
J.~Billings, A.~Malviya, and J.~Hendrik, ``{CW20: Jay Billings, Addi Malviya,
  John Hendrick Interview, RSE Group, Oak Ridge National Laboratory},''
  \url{https://youtu.be/WBnOLtac4B4}, accessed: 2020-08-07.

\bibitem{milewicz2020ebp}
R.~Milewicz, ``{Towards Evidence-Based Practice in Scientific Software
  Development},'' \emph{2020 Collegeville Workshop on Scientific Software},
  2020.

\bibitem{jay2017identifying}
C.~Jay, R.~Haines, M.~Vigo, N.~Matentzoglu, R.~Stevens, J.~Boyle, A.~Davies,
  C.~Del~Vescovo, N.~Gruel, A.~Le~Blanc \emph{et~al.}, ``{Identifying the
  challenges of code/theory translation: Report from the Code/Theory 2017
  Workshop},'' \emph{Research Ideas and Outcomes}, vol.~3, p. e13236, 2017.

\end{thebibliography}

\end{document}